\documentclass[aps,prd,epsecnum,preprint,nofootinbib]{revtex4}
\usepackage{bm}
\usepackage{graphicx}
\usepackage{color}

\newcommand{\bgamma}{\ensuremath{{\bar{\gamma}}}}

\begin{document}
\baselineskip5.5mm
\title{
Two-point correlation function of density perturbations in a large void universe
}
\author{
       ${}^{1,2}$Ryusuke Nishikawa \footnote{E-mail:ryusuke@sci.osaka-cu.ac.jp},
       ${}^{3}$Chul-Moon Yoo \footnote{E-mail:yoo@gravity.phys.nagoya-u.ac.jp},
and
        ${}^{1,4}$Ken-ichi Nakao \footnote{E-mail:knakao@sci.osaka-cu.ac.jp}
}
\affiliation{
${}^{1}$Department of Mathematics and Physics,
Graduate School of Science, Osaka City University,
3-3-138 Sugimoto, Sumiyoshi, Osaka 558-8585, Japan
\\
${}^{2}$APC (CNRS-Universit\'e Paris 7),
10 rue Alice Domon et L\'eonie Duquet, 75205 Paris Cedex 13, France
\\
${}^{3}$Division of Particle and Astrophysical Science, 
Graduate School of Science, Nagoya University, 
Furo-cho, Chikusa-ku, Nagoya 464-8602, Japan
\\
${}^{4}$DAMTP, Centre for Mathematical Sciences, University of Cambridge, Wilberforce
Road, Cambridge CB3 0WA, United Kingdom
}

\begin{abstract}
\baselineskip5.5mm
We study 
the two-point correlation function 
of density perturbations in a spherically symmetric void 
universe model which does not employ the Copernican principle. 
First we solve perturbation equations in the inhomogeneous universe model 
and obtain density fluctuations by using a method of non-linear perturbation theory 
which was adopted in our previous paper. 
From the obtained solutions, we calculate the two-point correlation function and
show that it has a local anisotropy at the off-center position
differently from those in homogeneous and isotropic universes.
This anisotropy is caused by the tidal force in the off-center region of the spherical void.
Since no tidal force exists in homogeneous and isotropic universes,
we may test the inhomogeneous universe by observing statistical distortion of 
the two-point galaxy correlation function.
\end{abstract}

\preprint{OCU-PHYS 383}

\preprint{AP-GR 106}

\maketitle

\section{introduction}\label{sec1}

Most of modern cosmological models are based on the Copernican principle 
which states the earth is not at a privileged position in the universe. 
The observed isotropy of the Cosmic Microwave Background (CMB) radiation 
together with the Copernican principle 
implies our universe is homogeneous and isotropic,
if the small scale structures less than 50 Mpc are coarse-grained.
Although the standard cosmology can explain a lot of observations naturally, 
we should note that the Copernican principle on cosmological scales $\geq$ 1 Gpc has not been confirmed. 
This means modern cosmology would contain systematic errors
that arise from the inhomogeneity of the background universe. 
The systematic errors may mislead us 
when we consider major issues in modern cosmology such as probing dark energy abundance 
and testing general relativity at cosmological scales. 
Thus, it is an unavoidable task in modern precision cosmology to test the Copernican principle. 

In order to test the Copernican principle, 
we have to investigate non-Copernican cosmological models which drop the Copernican principle. 
Non-Copernican models commonly assume that we live close to the center in a spherically symmetric spacetime 
since the universe is observed to be nearly isotropic around us. 
These models have also been studied as an alternative to dark energy, 
because some of them can explain the observation of Type Ia supernovae without introducing dark energy~\cite{Celerier:1999hp,Celerier:2009sv,Clifton:2008hv,Goodwin:1999ej,Iguchi:2001sq,Kolb:2009hn,Mustapha:1998jb,Tomita:1999qn,Tomita:2000jj,Tomita:2001gh,Vanderveld:2006rb,Yoo:2008su,Yoo:2010qn}. 
The non-Copernican models without dark energy have been tested by observations including the CMB acoustic peaks~\cite{Alexander:2007xx,Alnes:2005rw,Biswas:2010xm,Clarkson:2010ej,GarciaBellido:2008nz,Marra:2011ct,Marra:2010pg,Moss:2010jx,Nadathur:2010zm,Yoo:2010qy,Zibin:2008vk}, 
the kinematic Suniyaev-Zeldovich effect~\cite{Bull:2011wi,GarciaBellido:2008gd,Moss:2011ze,Yoo:2010ad,Zhang:2010fa} 
and others~\cite{Alnes:2006pf,Alnes:2006uk,Bolejko:2005fp,Clarkson:2012bg,dePutter:2012zx,Dunsby:2010ts,Enqvist:2009hn,Enqvist:2006cg,GarciaBellido:2008yq,Goto:2011ru,Kodama:2010gr,Mishra:2012vi,Quartin:2009xr,Regis:2010iq,Romano:2009mr,Romano:2010nc,Romano:2011mx,Tanimoto:2009mz,Uzan:2008qp,Yagi:2012vx,Yoo:2010hi,Zibin:2011ma,Zumalacarregui:2012pq},
and significant observational constraints exist.
However, it should be noted that even if we accept dark energy components, 
the existence of the large spherical inhomogeneity may significantly affects observational results
irrespective of the observational constraints
(see e.g. Ref.~\cite{Valkenburg:2013qwa}).
A large void universe which assumes we live at a center of a huge void  
whose radius is larger than 1 Gpc is known as one of popular models among the non-Copernican cosmologies,
and we take such model into consideration in this paper, although our strategy is rather general.

Growth of the large-scale structure in the universe 
can be thought as one of the 
most useful tools to examine the large spherical void model,
because the evolution of perturbations is expected to reflect the tidal force in the background spacetime.
Unfortunately, linear perturbation equations in the spherical void universe have not been solved~\cite{Gerlach:1980},
because the number of isometries in a spherically symmetric inhomogeneous spacetim 
are less than in a homogeneous and isotropic spacetime.
Though some authors~\cite{Alonso:2012ds,Alonso:2010zv,Clarkson:2009sc,February:2012fp,Zibin:2008vj} 
have studied the perturbation equations 
using a local-Friedmann-Lema\^{\i}tre-Robertson-Walker (FLRW) approximation 
which neglects shear of the background spacetime, 
it is not clear how to evaluate the accuracy for the approximation. 
Actually, in this paper,
we will show that the shear effect plays an important role in the growth of the perturbations
by using another complementary analytic approach 
proposed in our last paper~\cite{Nishikawa:2012we}.

In our previous work~\cite{Nishikawa:2012we}, 
we considered two kinds of perturbations in the homogeneous and isotropic universe. 
One is the isotropic mode which represents 
the large spherical void, 
and the other is the anisotropic mode which denotes the large-scale structure besides the void. 
We solved successively nonlinear perturbation equations in the homogeneous and isotropic universe model, 
where the isotropic and anisotropic perturbations couple with each other, 
and then the evolution of anisotropic density fluctuations affected by the spherical void was clarified.
By using the solution for the non-linear perturbation equations,
we calculated angular power spectrum which is defined as the two-point 
correlation of the density perturbations 
in the direction transverse to the line of sight of the observer at the center of the void.
By computing the growth rate of the angular power spectrum, 
we showed the growth of perturbations in the void universe model 
is different from those in the homogeneous and isotropic universes.
However, we have studied the correlation of the density perturbations 
only for the direction transverse to the line of sight of the central observer.
In this paper, we will calculate the two-point correlation function 
of the density perturbations in all directions 
and discuss the direction-dependence, that is, distortion of the two-point correlation function
which is caused by the shear of the huge void in an off-center region.

This paper is organized as follows. 
In \S~\ref{sec2}, we review our method for solving perturbation equations. 
In \S~\ref{sec3}, we calculate the two-point correlation function 
of the density fluctuations, and discuss its distortion.
In \S~\ref{sec4}, we evaluate the 
distortion of the two-point correlation function in simple void models. 
 \S~\ref{sec5} is devoted to a summary and discussion.

In this paper, we use the geometrized units in which 
the speed of light and Newton's gravitational constant are one, respectively. 
The Latin indices denote the spatial components, whereas the Greek indices represent
the spacetime components.

\section{perturbations in a large void universe}\label{sec2}

We choose the dust-$\Lambda$-FLRW universe model as a background, 
whose metric and stress energy tensor is given by
\begin{eqnarray}
 ds^2&=&-dt^2+a^2(t)\bgamma_{ij}dx^idx^j  \cr
 && \cr
 &:=& -dt^2+a^2(t)\left[d\chi^2+S_K^2(\chi)(d\theta^2+\sin^2\theta d\phi^2)\right],
\end{eqnarray}
and
\begin{eqnarray}
 T^{\mu \nu }&=&\bar{\rho }(t)\bar{u}^\mu \bar{u}^\nu ,
\end{eqnarray}
where $\bar{\rho}$, $\bar{u}^\mu=(1,0,0,0) $ and $a(t)$ are 
the energy density, 4-velocity of dust fluid element and scale factor, respectively, 
and $S_K$ is defined by
\begin{eqnarray}
 S_K(\chi)
 &=&
 \left\{
 \begin{array}{ll}
 \sinh\left(\sqrt{-K}\chi\right)/\sqrt{-K}, & \mbox{$K<0$} \\
 \chi, & \mbox{$K=0$} \\
 \sin\left(\sqrt{K}\chi\right)/\sqrt{K}, & \mbox{$K>0$}
 \end{array}
 \right.
 \nonumber
\end{eqnarray}
where $K$ is a constant that denotes spatial curvature.

We consider perturbations parametrized by two book-keeping parameters $\kappa$ and $\epsilon$ 
on the dust-$\Lambda$-FLRW universe shown above as a background; 
both $\kappa$ and $\epsilon$ represent the smallness of perturbations 
but with $0<\epsilon\ll\kappa\ll1$ during calculations. 
The fluctuations with $\kappa$ are isotropic and compose a spherical void inhomogeneity, 
whereas the perturbations with $\epsilon$ are anisotropic
and denote the large-scale structures such as clusters of galaxies.

In the synchronous comoving gauge, the metric and stress-energy tensor of the perturbed universe model is written as
\begin{eqnarray}
 ds^2
 &=&
 -dt^2+a^2(t)\sum_{N=0}\kappa^N
 \left[\ell^{(N)}_{ij}dx^idx^j +\epsilon \; h^{(N+1)}_{ij}dx^idx^j +{\cal O}(\epsilon^2)\right],
\label{ds:1}
\end{eqnarray}
and 
\begin{eqnarray}
 T^{\mu \nu }
 &=&
 \bar{\rho }(t)\bar{u}^\mu \bar{u}^\nu 
 \sum_{N=0}\kappa^N\left[\Delta ^{(N)}(t,\chi)+\epsilon \; \delta ^{(N+1)}(t,{\bf x})+{\cal O}(\epsilon^2)\right],
 \label{tmn:1}
\end{eqnarray}
where $\ell^{(0)}_{ij}=\bgamma_{ij}$, $\Delta^{(0)}=1$ and ${\bf x}:=(\chi,\theta,\phi)$.
Here, we note two limits, $\epsilon\rightarrow0$ with $\kappa\neq0$ and $\kappa\rightarrow0$ with $\epsilon\neq0$. 
In the former case, the spacetime coincides with the $\Lambda$-Lema\^{\i}tre-Tolman-Bondi ($\Lambda$-LTB) solution 
which is the spherically symmetric dust solution of the Einstein equation with the cosmological constant $\Lambda$, 
if we take all orders of the $\kappa$ into account.
In the later case, the spacetime coincides with the homogeneous and isotropic universe 
with standard linear perturbations, if we neglect the terms of the order higher than or equal to $\epsilon^2$.

By substituting expressions (\ref{ds:1}) and (\ref{tmn:1}) 
into the Einstein equations $G_{\mu\nu}=8\pi T_{\mu\nu}$ 
and the equation of motion for matter $\nabla_\mu T^{\mu\nu}=0$,
and assuming the equations hold in each order with respect to $\kappa$ and $\epsilon$, 
we obtain the equations for the density perturbations of the order $\epsilon$ and $\kappa \epsilon$ as follows;
\begin{eqnarray}
 \ddot{\delta}^{(1)}+2H\dot{\delta}^{(1)}-4\pi \bar{\rho }\delta^{(1)}&=&0,
 \label{per:1} \\
 \ddot{\delta}^{(2)}+2H\dot{\delta}^{(2)}-4\pi \bar{\rho }\delta^{(2)}&=&S^{(2)}, 
 \label{per:2}
\end{eqnarray}
where a dot denotes a time derivative and $H:=\dot{a}/a$, 
and by denoting $\ell^{ij}:=\bgamma^{ik}\bgamma^{jl}\ell_{kl}$, 
\begin{eqnarray}
 S^{(2)}
 &:=&
 \frac{1}{2}\dot{\ell }^{(1)ij}\dot{h}^{(1)}_{ij} 
 +2\dot{\Delta}^{(1)}\dot{\delta }^{(1)}
 +\ddot{\Delta}^{(1)}\delta ^{(1)}
 +\Delta ^{(1)}\ddot{\delta}^{(1)}+2H\dot{\Delta}^{(1)}\delta^{(1)} +2H\Delta^{(1)} \dot{\delta }^{(1)}.
\end{eqnarray}
The general solution of Eq.~(\ref{per:1}) is represented 
by a linear superposition of the growing factor $D^+(t)$ and decaying factor $D^-(t)$ which are given by
\begin{eqnarray}
 D^+(t)=H\int^{a(t)}\frac{da}{a^3H^3}~~~{\rm and}~~~D^-(t)=H.
\end{eqnarray}
Here for simplicity, we ignore the decaying mode, 
and $\delta^{(1)}(t,{\bf x})=D^+(t)\delta^{(1)} (t_{\rm i},{\bf x})$ 
where $t_{\rm i}$ is sufficiently early time so that
the isotropic perturbations are negligible 
and standard homogeneous and isotropic cosmology is applicable at this stage.
We solve Eq.~(\ref{per:2}) by using the Green function method and obtain 
\begin{eqnarray}
 \delta (t,{\bf x})
 &:=&
 \epsilon \; \delta^{(1)}(t,{\bf x})+\kappa \epsilon \; \delta^{(2)}(t,{\bf x}) \cr
 &&\cr
 &=&
 \epsilon \; \delta^{(1)}(t,{\bf x})
 +\kappa \epsilon \left[T_1(t)\Delta^{(1)}(t_{\rm i},\chi)\delta^{(1)}(t,{\bf x})
 +T_2(t)\dot{\ell}^{(1)ij}(t_{\rm i},{\bf x})h_{ij}^{(1)}(t,{\bf x})\right],
 \label{per:3}
\end{eqnarray}
where
\begin{eqnarray}
 T_1(t)&:=&[D^+(t)]^{-1}\int ^t_{t_{\rm i}}dsG(s;t)
 \left(
 2\dot{D}^+(s)\dot{D}^+(s)+2\ddot{D}^+(s)D^+(s)+4H(s)\dot{D}^+(s)D^+(s)
 \right),
 \nonumber \\
 T_2(t)&:=&\frac{1}{2}[D^+(t)\dot{D}^+(t_{\rm i})]^{-1}\int ^t_{t_{\rm i}}dsG(s;t)\dot{D}^+(s)\dot{D}^+(s),
 \nonumber \\
 G(s;t)&:=&\frac{D^-(t)D^+(s)-D^+(t)D^-(s)}{D^+(s)\dot{D}^-(s)-\dot{D}^+(s)D^-(s)}.\nonumber
\end{eqnarray}

For later convenience, focusing on the scalar modes on a sphere specified by the radial coordinate $\chi$, 
we rewrite the term $\dot{\ell}^{(1)ij}h^{(1)}_{ij}$ in the solution (\ref{per:3}) as follows.
We first rewrite the metric perturbations of the order $\kappa$ in the form
\begin{eqnarray}
 \ell^{(1)}_{ij}dx^idx^j
 =
 \ell^{(1)}_{||}(t,\chi)d\chi^2+\ell^{(1)}_\bot (t,\chi)S_K^2(\chi)d\Omega^2.
\end{eqnarray}
Note that $\ell_{||}^{(1)}(t,\chi)$ and $\ell^{(1)}_\bot (t,\chi)$ are the scalar on a sphere specified by $\chi$. 
Then, we define the radial and azimuthal Hubble parameters as
\begin{eqnarray}
 H_{||}:=H+\kappa \psi^{(1)}_{||}+{\cal O}(\epsilon,\kappa^2)~~~~{\rm and}~~~~~
 H_{\bot}:=H+\kappa \psi^{(1)}_{\bot}+{\cal O}(\epsilon, \kappa^2),
\end{eqnarray}
where
\begin{eqnarray}
 \psi_{||}^{(1)}:=\frac{1}{2}\dot\ell_{||}^{(1)}~~~~{\rm and}~~~~
 \psi_{\bot}^{(1)}:=\frac{1}{2}\dot\ell_{\bot}^{(1)}.
\end{eqnarray}
Regarding the anisotropic perturbations, 
we assume that the metric perturbation $h_{ij}^{(1)}$ is composed of the only scalar modes, 
and thus it is written in the form
\begin{eqnarray}
 h_{ij}^{(1)}=\Psi^{(1)}(t,{\bf x})\bgamma_{ij}+\mathcal{D}_i\mathcal{D}_j\Phi^{(1)}(t,{\bf x}),
\end{eqnarray}
where $\mathcal{D}_i$ denotes covariant derivative with respect to $\bgamma_{ij}$.  
Note that both $\Psi^{(1)}$ and $\Phi^{(1)}$ are the scalar on the hypersurface of constant $t$. 
The time-space components of the Einstein equations of the order $\epsilon$ lead to $\Psi^{(1)}=K\Phi^{(1)}$, 
and hence we have
\begin{eqnarray}
 h_{ij}^{(1)}=K\Phi^{(1)}(t,{\bf x})\gamma_{ij}+\mathcal{D}_i\mathcal{D}_j\Phi^{(1)}(t,{\bf x}).
\end{eqnarray}
The function $\Phi^{(1)}$ is related to the density contrast $\delta^{(1)}$ 
through the equation of motion for the dust of the order $\epsilon$ as
\begin{eqnarray}
 \delta^{(1)}&=&-\frac{1}{2}\left(\mathcal{D}^i\mathcal{D}_i+3K\right)\Phi^{(1)}.
\label{P-eq}
\end{eqnarray}
By using the functions $\psi_{||}^{(1)}, \psi_\bot^{(1)}$ and $\Phi^{(1)}$, 
the density contrast (\ref{per:3}) is reduced to
\begin{eqnarray}
 \delta (t,{\bf x})&=&\epsilon \; \delta^{(1)}(t,{\bf x})
 +\kappa \epsilon \biggl[T_1(t)\Delta^{\rm i}(\chi)\delta^{(1)}(t,{\bf x}) \cr 
 &+&2T_2(t)\left\{
 K\left(\psi^{\rm i}_{||}(\chi)+2\psi_\bot ^{\rm i}(\chi)\right)+\psi_\bot ^{\rm i}(\chi)\mathcal{D}^i\mathcal{D}_i
 +\left(\psi_{||}^{\rm i}(\chi)-\psi_\bot^{\rm i}(\chi)\right)\partial_\chi^2\right\}\Phi^{(1)}(t,{\bf x}) \biggr]\cr
 &&\cr
 &=&\epsilon \; \delta^{(1)}(t,{\bf x})
 +\kappa \epsilon \biggl[\left\{T_1(t)\Delta^{\rm i}(\chi)-4T_2(t)\psi_\bot^{\rm i}(\chi)\right\}
 \delta^{(1)}(t,{\bf x}) \cr
 &+&2T_2(t)\left\{\psi_{||}^{\rm i}(\chi)-\psi_\bot^{\rm i}(\chi)\right\}
 \left(K+\partial_\chi^2\right)\Phi^{(1)}(t,{\bf x}) \biggr],
\label{dens:4}
\end{eqnarray}
where the superscript ${\rm i}$ represents the initial value at $t=t_{\rm i}$,
and we have used Eq.~(\ref{P-eq}) in the second equality.

\section{derivation of two-point correlation function}\label{sec3}

\begin{figure}[htbp]
 \begin{center}
 \includegraphics[width=6cm,clip]{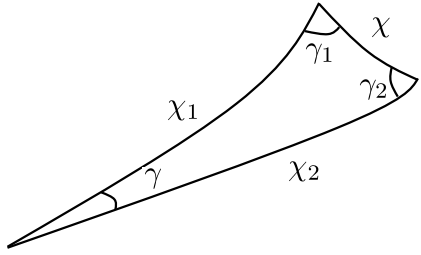}
 \end{center}
 \caption{
 Geometry of the relative position of the observer and two points 
 in the constant curvature space~\cite{Matsubara:1999du}.
 }
 \label{tri}
\end{figure}

As already mentioned,
it is the purpose of this paper to study two-point correlation function 
of density perturbations in inhomogeneous and isotropic universe models. 
In order to clarify the evolution of density perturbations, we have invoked
the perturbative analysis on the background dust-$\Lambda$-FLRW universe.
By virtue of this treatment, we can specify the relative position of two points and the central observer 
by using the {\it comoving distance} 
which is the geodesic distance with respect to the background conformal metric $\bgamma_{ij}$.
We represent two-point correlation functions of anisotropic density perturbations 
in the inhomogeneous and isotropic universe model in the form 
\begin{eqnarray}
 \xi (t,{\bf x}_1,{\bf x}_2):=\langle \delta (t,{\bf x}_1)\delta (t,{\bf x}_2)\rangle 
 =\xi (t,\chi,\chi_1,\chi_2),
 \label{cf0}
\end{eqnarray}
where $\chi_{1,2}$ are the comoving distances from the central observer to the points, 
and $\chi$ is the comoving separation of the two points (see fig.~\ref{tri}).

By using the quantities introduced in the previous section, the two-point correlation function is given by
\begin{eqnarray}
 \xi &=&\epsilon^2\langle \delta^{(1)} (t,{\bf x}_1)\delta^{(1)} (t,{\bf x}_2)\rangle \cr
 &&\cr
 &+&\kappa \epsilon^2\langle \delta^{(2)} (t,{\bf x}_1)\delta^{(1)} (t,{\bf x}_2)\rangle 
 +\kappa \epsilon^2\langle \delta^{(1)} (t,{\bf x}_1)\delta^{(2)} (t,{\bf x}_2)\rangle 
+{\cal O}(\epsilon^2\kappa^2).
\label{xi}
\end{eqnarray}
The terms of the order $\kappa\epsilon^2$ in the right hand side of the above equation 
represent the effects of the spherical void on the anisotropic perturbations. 
By using Eq.~(\ref{dens:4}), these terms of our interest are written as follows. 
The second term in the right hand side of Eq.~(\ref{xi}) is given by
\begin{eqnarray}
 &&\langle\delta^{(2)}(t,{\bf x}_1)\delta^{(1)}(t,{\bf x}_2)\rangle 
 =\left[T_1(t)\Delta^{\rm i}(\chi_1)-4T_2(t)\psi^{\rm i}_\bot(\chi_1)\right]
 \langle\delta^{(1)}(t,{\bf x}_1)\delta^{(1)}(t,{\bf x}_2)\rangle \cr
 && \cr
 &&~~~~~~~~~~~~+2T_2(t)\left[\psi_{||}^{\rm i}(\chi_1)-\psi_\bot^{\rm i}(\chi_1)\right]
 \left(K+\partial_{\chi_1}^2  \right)
 \langle\Phi^{(1)} (t,{\bf x}_1)\delta^{(1)} (t,{\bf x}_2)\rangle.
 \label{col:1}
\end{eqnarray}
Hereafter, we assume that the wavelength $\lambda$ of the anisotropic perturbations 
is much smaller than the scale of the spatial curvature, $\partial_\chi\sim1/\lambda\gg \sqrt{|K|}$, 
and thus we discard the term proportional to $K$ in the above equation.
\footnote{We note that $\chi_1$ and $\chi_2$ can be the same order of $1/\sqrt{|K|}$.}
The third term in the right hand side of Eq.~(\ref{xi}) 
is obtained by replacing the subscript 1 by 2 and 2 by 1, 
except for the subscript of $T_1$ and $T_2$, in Eq.~(\ref{col:1}). 

Under the short-wavelength assumption, $\lambda\ll 1/\sqrt{|K|}$, 
the two-point correlation function of the linear density perturbations $\delta^{(1)}$ 
is written as~\footnote{General formula which does not employ the short-wavelength approximation 
can be seen in Matsubara's paper~\cite{Matsubara:1999du}.}
\begin{eqnarray}
 \langle\delta^{(1)}(t,{\bf x}_1)\delta^{(1)}(t,{\bf x}_2)\rangle &=&
 \int_0^\infty \frac{dkk^2}{2\pi^2}j_0(k\chi)P^{(1)}(t,k),
 \label{powe:1}
\end{eqnarray}
where $P^{(1)}(t,k)$ is the so-called power spectrum in the homogeneous and isotropic universes.
By using Eq.~(\ref{powe:1}) 
together with the relation
$\delta^{(1)}\simeq -\frac{1}{2}\mathcal{D}^i\mathcal{D}_i \Phi^{(1)}$ from Eq.~(\ref{P-eq}), we obtain
\begin{eqnarray}
 \langle\Phi^{(1)}(t,{\bf x}_1)\delta^{(1)} (t,{\bf x}_2)\rangle =
 2\int \frac{dk k^2}{2\pi^2}j_0(k\chi)\frac{P^{(1)}(t,k)}{k^2}.
 \label{Phi:2}
\end{eqnarray}
The remaining nontrivial term of the equation (\ref{col:1}) is the derivative with respect to $\chi_1$.  
To evaluate the term of the derivative with respect to $\chi_1$ in Eq.~(\ref{col:1}),
we use the following useful formula (see Appendix~\ref{secap1} and Ref.~\cite{Matsubara:1999du} for the derivation)
\begin{eqnarray}
 \frac{\partial^2j_0(k\chi)}{\partial \chi_1^2}=-\frac{k^2}{3}j_0(k\chi)+\frac{k^2}{3}(3\cos^2\gamma_1-1)j_2(k\chi).
 \label{cos:2}
\end{eqnarray}
By using Eqs.~(\ref{col:1}), (\ref{powe:1}), (\ref{Phi:2}) and (\ref{cos:2}), we finally obtain
\begin{eqnarray}
 &&\xi (t,\chi,\chi_1,\chi_2)=\epsilon^2\xi_{(0)}(t,\chi) \cr
 &&\cr
 &&~~~~~~~~+\kappa \epsilon^2\left[A(t,\chi_1,\chi_2) \xi_{(0)}(t,\chi)
 +B(t,\chi,\chi_1,\chi_2) \xi_{(2)}(t,\chi)\right] +{\cal O}(\kappa^2\epsilon^2),
 \label{cf1}
\end{eqnarray}
where $\xi_{(l)}$ is defined by
\begin{eqnarray}
 \xi_{(l)}(t,\chi):=\int_0^\infty \frac{dk}{2\pi^2}k^2j_l(k\chi)P^{(1)}(t,k), \nonumber
\end{eqnarray}
and, by using the Legendre polynomial of degree two, $P_2(z)$,
\begin{eqnarray}
 A(t,\chi_1,\chi_2)
 &:=&T_1(t)\left[\Delta^{\rm i}(\chi_1)+\Delta^{\rm i}(\chi_2)\right] \cr
 && \cr
 &&-\frac{4}{3}T_2(t)\left[\psi^{\rm i}_{||}(\chi_1)+2\psi^{\rm i}_\bot(\chi_1)
 +\psi^{\rm i}_{||}(\chi_2)+2\psi^{\rm i}_\bot(\chi_2) \right], \cr
 && \cr
 B(t,\chi,\chi_1,\chi_2)
 &:=&\frac{8}{3}T_2(t)\biggl[P_2(\cos \gamma_1) \left\{\psi^{\rm i}_{||}(\chi_1)-\psi^{\rm i}_\bot(\chi_1)\right\} \cr
 &&\cr
 &&+P_2(\cos \gamma_2)\left\{\psi^{\rm i}_{||}(\chi_2)-\psi^{\rm i}_\bot(\chi_2)\right\}
 \biggr], \nonumber
\end{eqnarray}
where $\gamma_{1,2}$ in the above equation are represented by $\chi_1, \chi_2$ and $\chi$
(see Eqs.~(\ref{K-})--(\ref{K+}) in Appendix~\ref{secap1}).

So far, we have only assumed $0<\kappa\ll1$ and $\partial_\chi\gg \sqrt{|K|}$.
The spatial curvature $K$ and the cosmological constant $\Lambda$ have not been ignored, 
and further any specific spatial configurations for the isotropic perturbations have not been assumed yet. 
Therefore, the equation~(\ref{cf1}) can be used to wide class of inhomogeneous and isotropic universes.

To clarify the behavior of the two-point correlation function~(\ref{cf1}) in the model of huge void universe, 
we focus on the following situation; 
we consider two-point correlations
whose comoving separation $\chi$ is much smaller than both the comoving scale of the void $L_{\rm void}$ 
and the comoving distance from the central observer to these points, $\chi_1$ and $\chi_2$.  
By the first assumption, $\chi\ll L_{\rm void}$, we have 
\begin{eqnarray}
 \Delta^{(1)} (t,\chi_2)
 &=&\Delta^{(1)} (t,\chi_1)+\sum_{n=1}\frac{1}{n!}(\chi_2-\chi_1)^n\frac{\partial^n 
 \Delta^{(1)}(t,y)}{\partial y^n}\biggl|_{y=\chi_1}  \cr
 &&\cr
 &=& \left[1+\mathcal{O}\left(\frac{\chi}{L^{\rm void}}\right) \right]\Delta^{(1)} (t,\chi_1),
\end{eqnarray}
where we have used $\chi\leq |\chi_1-\chi_2|$ for the second equality. 
The similar relations as the above also hold for $\psi^{(1)}_{||}$ and $\psi^{(1)}_\bot$. 
Then, by the second assumption, $\chi \ll \chi_{1,2}$,
which is often called the distant-observer approximation~\cite{Matsubara:1996nf},
the angle $\gamma_2$ can be approximated as $\gamma_2\simeq \pi-\gamma_1$.
By these two assumptions, the two-point correlation function~(\ref{cf1}) is reduced to
\begin{eqnarray}
 \xi(t,\chi,\chi_1,\chi_2)&\simeq& \xi_{\rm ap}(t,\chi,\chi_1,\gamma_1) \cr
 &&\cr
 &:=&\epsilon^2\xi_{(0)}(t,\chi)
 +\kappa \epsilon^2\left[a(t,\chi_1)\xi_{(0)}(t,\chi)+b(t,\chi_1,\gamma_1)\xi_{(2)}(t,\chi)
 \right],
 \label{cf2}
\end{eqnarray}
where
\begin{eqnarray}
 a(t,\chi_1)
 &:=&2T_1(t)\Delta^{\rm i}(\chi_1)-\frac{8}{3}T_2(t)
 \left(\psi^{\rm i}_{||}(\chi_1)+2\psi^{\rm i}_\bot (\chi_1)\right), \cr
 && \cr
 b(t,\chi_1,\gamma_1)
 &:=&\frac{16}{3}T_2(t)P_2(\cos \gamma_1) \left[\psi^{\rm i}_{||}(\chi_1)-\psi^{\rm i}_\bot(\chi_1)\right]. 
 \nonumber
\end{eqnarray}
In the the above equations, the $\chi_1$-dependence implies the inhomogeneity of the two-point 
correlation function,
which comes from the spherical perturbations, $\Delta^{(1)}, \psi^{(1)}_{||}$ and $\psi^{(1)}_\bot$.
We can also see that the $\gamma_1$-dependence of $b(t,\chi_1,\gamma_1)$ 
corresponds to the distortions of the correlation, 
which results from the local anisotropy of the volume expansion rate at $\chi_1\neq0$, 
that is, $\psi^{(1)}_{||}-\psi^{(1)}_\bot\neq0 $.
We would like to stress that the local-FLRW approximation never predicts the existence of a term 
that represents the $\gamma_1$-dependence of the two-point correlation function.
Since the function $T_2(t)$ is the growth factor of the second-order perturbations, 
the distortion of the correlation becomes important at late time.

By investigating the difference between its value of $\gamma_1=\pi$ and of $\gamma_1=\pi/2$, 
we can see whether distortion of the two-point correlation function exists. 
Here, it should be noted that if we take the distance up to the order $\kappa$, 
the comoving distance $\chi$ does not mean the same proper distance for $\gamma_1=\pi$ and $\gamma_1=\pi/2$.
By taking this fact into account, we define the following quantity 
\begin{equation}
\Pi (t,\chi_{\rm p},\chi_1):=\xi_{\rm ap}\left(t,\chi_\bot,\chi_1,\pi/2\right)
 -\xi_{\rm ap}\left(t,\chi_{||},\chi_1,\pi\right),
\label{pi1}
\end{equation}
where $\chi_{||}$ and $\chi_\bot$ are related to the proper distance $\chi_{\rm p}$ as
\begin{eqnarray}
 \chi_{||}=
 \chi_{\rm p}\left[1-\frac{\kappa}{2}\ell_{||}(t,\chi_1)\right]~~~{\rm and}~~~
 \chi_\bot =
 \chi_{\rm p}\left[1-\frac{\kappa}{2}\ell_\bot(t,\chi_1)\right].
\end{eqnarray}
Substituting Eq.~(\ref{cf2}) into Eq.~(\ref{pi1}), we have
\begin{eqnarray}
 \Pi (t,\chi_{\rm p},\chi_1)&\simeq& \kappa\epsilon^2
 \Biggl[
 \frac{\chi_{\rm p}}{2}\left\{\ell_{||}(t,\chi_1)-\ell_\bot (t,\chi_1)\right\}
 \frac{\partial \xi_{(0)}(t,\chi)}{\partial\chi}\biggr|_{\chi=\chi_{\rm p}}
 \cr
 &+&8T_2(t)\left\{\psi_\bot^{\rm i}(\chi_1)-\psi_{||} ^{\rm i}(\chi_1)\right\}\xi_{(2)}(t,\chi_{\rm p})
 \Biggr].
 \label{aniso}
\end{eqnarray}
The quantity $\Pi$ is a measure of the distortion of the two-point correlation function at each point. 

\section{Distortion of two-point correlation function}\label{sec4}

\begin{figure}[htbp]
 \begin{center}
 \includegraphics[width=8cm,clip]{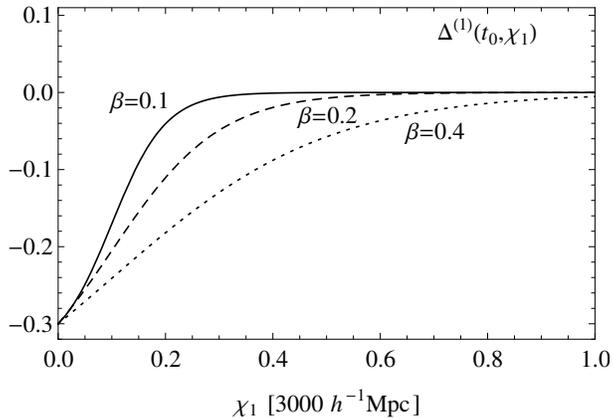}
 \end{center}
 \caption{
 The density contrast $\Delta^{(1)}$ at present time as a function of $\chi_1$ for 
 three cases, $\beta=0.1,~0.2$ and $0.4$.
 }
 \label{del}
\end{figure}

We investigate the distortion of the two-point correlation function $\xi$ in a specific model of the void universe.
We assume that this model approaches to the Einstein-de Sitter universe model in the spatial asymptotic region
with the dimensionless Hubble parameter $h:=H_0/100{\rm km s^{-1} Mpc^{-1}} =0.7$.
In the perturbative treatment, the inhomogeneity of the void model is described by 
the isotropic perturbations of the order $\kappa$ on the Einstein-de Sitter universe.
Since we consider the void model which can be approximated 
by the homogeneous and isotropic universe at early stage, 
we neglect the decaying mode for the perturbations of the order $\kappa$.
We fix the gauge degree of freedom to rescale the radial coordinate 
as $\ell_\bot^{(1)} (t_0,\chi)=0$, where $t_0$ is present time.
Then the isotropic perturbations are completely determined 
by the growing mode $\Delta^+(\chi)$, where the density contrast is given by 
\begin{eqnarray}
 \Delta^{(1)}(t,\chi)=\frac{D^+(t)}{D^+(t_0)}\Delta^+(\chi).
\end{eqnarray}
We present calculations to determine other perturbations, $\ell_{||}^{(1)}$ and $\ell_\bot^{(1)},$ from $\Delta^+$
in Appendix~\ref{secap2}.
We set the function $\Delta^+$ as
\begin{eqnarray}
 \Delta^+(\chi_1)&=&
 -0.3\times\frac{1-\tanh\left[(\chi_1-0.1)/\beta\right]}{1+\tanh\left[0.1/\beta\right]},
 \label{conf:1}
\end{eqnarray}
where $\beta$ is a parameter that determines the size of void.
We set the amplitude of the isotropic density perturbation to be about 0.3 at present time. 
We show the density contrast at present time, $\Delta^{(1)}(t_0,\chi_1)$, 
for three cases, $\beta=0.1,~0.2$ and $0.4$, as functions of $\chi_1$ in Fig.~\ref{del} .
We can see from this figure that the size of the void is about
$750h^{-1}{\rm Mpc}$ for $\beta=0.1$, $1500h^{-1}{\rm Mpc}$ for $\beta=0.2$ 
and $3000 h^{-1}{\rm Mpc}$ for $\beta=0.4$, respectively.

\begin{figure}[htbp]
 \begin{center}
 \includegraphics[width=8cm,clip]{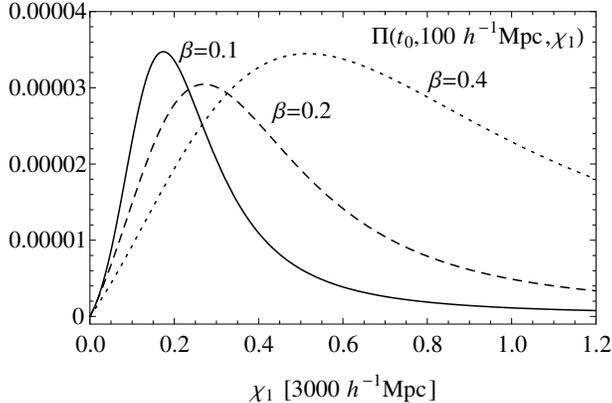}
 \end{center}
 \caption{
 The quantity $\Pi$ which represents 
 the distortions of the two-point correlation function for $\chi_{\rm p} =100 h^{-1}$ Mpc at the present time $t=t_0$ 
 as a function of $\chi_1$ for cases, $\beta=0.1,~0.2$ and $0.4$.
 }
 \label{pow1}
\end{figure}

We depict the quantity $\Pi$ at the present time $t=t_0$ as a function of $\chi_1$  for three cases, 
$\beta=0.1,~0.2$ and 0.4 in Fig.~\ref{pow1}. 
Here, we have chosen the proper distance $\chi_{\rm p}$ between two points to be equal to $100h^{-1}$ Mpc,
and we have used the fitting formula for the power spectrum $P^{(1)}(t,k)$
developed by Eisenstein \& Hu~\cite{Eisenstein:1997ik}.
We can see from Fig.~\ref{pow1} that the maximum of $\Pi$ is located near the edge of the void. 
It is worth to notice that the magnitude of the two-point correlation function of the order $\epsilon^2$ is 
$\xi_{(0)}(t_0,100 h^{-1} {\rm Mpc})\simeq -1.2\times 10^{-4}$. 
Then, we can also see from Fig.~\ref{pow1} that the function $\Pi$ is about 
quarter of the leading order term of the two-point correlation function.   
So, we conclude that 
the distortion of the two-point correlation function is important in observationally
studying the growth of the large-scale structure in a large void universe.

\section{summary and discussion}\label{sec5}
We have derived an expression for two-point correlation function of density perturbations 
in the inhomogeneous and isotropic model of the universe,
by applying the second-order perturbation theory in the homogeneous and isotropic universe.
First, we have derived the general expression~(\ref{cf1}) for the two-point correlation function
in a spherical inhomogeneous universe model in a form of the series expansion.
Then, we have assumed the separation between two points which we take the correlation is much shorter than 
both the scale of the spherical inhomogeneity and the distance from the center.
In these approximation, it can be explicitly shown that
the two-point correlation function has the distortion as a result of the local anisotropy of the volume expansion rate. 
This result is very different from the prediction based on the so-called local-FLRW approximation 
in which sufficiently small region is assumed to be the same as the FLRW universe. 
Our result suggests that we should treat a large void universe model 
as a locally homogeneous and anisotropic universe model rather than a locally FLRW universe model.

We computed the distortion of the two-point correlation function for a specific model
with the order of the spherical inhomogeneity being about 10\%. 
In this model, the magnitudes of the distortions are not negligible 
compared to the leading order term in the two-point correlation function. 
Hence, we may test the model of the huge void universe 
by the observations of the distortion of the two-point correlation function. 
In other words, the observational data of the two-point correlation function of galaxy distribution 
may contain a systematic error due to the non-Copernican inhomogeneity.

In practice, the galaxy distribution is observed not in the real space but in the redshift space. 
In the case of the homogeneous and isotropic universe model, 
it is known that coherent peculiar velocity of the galaxies leads to redshift distortions
in the clustering pattern of galaxies in redshift space. 
Recently, Guzzo et al.~\cite{Guzzo:2008ac} and Blake et al.~\cite{Blake:2011rj} have presented 
the observational results on the distortions of the power spectra 
which is consistent with the prediction in the $\Lambda$CDM model. 
So, if the distortions with non-Copernican inhomogeneity are significantly different from 
that in the $\Lambda$CDM model, 
we may give a significant constraint for the non-Copernican inhomogeneity
using these observational results. 
In the models of the spherical void universe,
we expect that the distortion of the power spectra comes from 
both the tidal force of the void and the peculiar velocity of galaxies.
The effect of the peculiar velocity to the redshift distortion in the void model is left for future work.
In order to compare our theoretical prediction with the observational data,
we need to obtain the two-point correlation function in the redshift space. This is also left for future work.

\section*{Acknowledgments}

RN is supported by a Grant-in-Aid through the Japan Society for the Promotion of Science (JSPS).
RN is also supported by the JSPS Strategic Young Researcher
Overseas Visits Program for Accelerating Brain Circulation
``Deepening and Evolution of Mathematics and Physics,
Building of International Network Hub based on OCAMI''. 
This work was supported in part by JSPS Grant-in-Aid for Scientic Research (C) (No. 25400265)

\appendix

\section{Derivation of Eq.~(\ref{cos:2})}
\label{secap1}
We use the following relations 
(see, for example~\cite{Matsubara:1999du}); for $K<0$,
\begin{equation}
\cosh(\sqrt{-K}\chi)=\cosh(\sqrt{-K}\chi_1)\cosh(\sqrt{-K}\chi_2)
-\sinh(\sqrt{-K}\chi_1)\sinh(\sqrt{-K}\chi_2)\cos\gamma; \label{K-}
\end{equation}
for $K=0$, 
\begin{equation}
\chi^2=\chi_1^2+\chi_2^2-2\chi_1\chi_2\cos\gamma; 
\end{equation}
for $K>0$, 
\begin{equation}
\cos(\sqrt{K}\chi)=\cos(\sqrt{K}\chi_1)\cos(\sqrt{K}\chi_2)+\sin(\sqrt{K}\chi_1)\sin(\sqrt{K}\chi_2)\cos\gamma. 
\label{K+}
\end{equation}
Furthermore, the following relations hold
$$
\cos\gamma_1=\frac{\partial\chi}{\partial\chi_1}~~~~~{\rm and}~~~~~
\cos\gamma_2=\frac{\partial\chi}{\partial\chi_2},
$$
where $\gamma_1$ ($\gamma_2$) is defined as an angle between the geodesics of $\chi_1$ ($\chi_2$) 
and $\chi$ (see fig.~\ref{tri}).
By differentiating Eqs.~(\ref{K-})--(\ref{K+}) with respect to $\chi_1$ with $\chi_2$ and $\gamma$ fixed, 
we obtain $\partial \chi/\partial\chi_1$ and $\partial^2\chi/\partial\chi_1^2$. 
Then, by using these results, we obtain Eq.~(\ref{cos:2}). 

\section{Perturbations of the order $\kappa$ on the Einstein-de Sitter model}
\label{secap2}

Perturbation equations of the order $\kappa$ on the Einstein-de Sitter model are written as
\begin{eqnarray}
 \ddot{\Delta}^{(1)}+2H\dot{\Delta}^{(1)}-4\pi \bar{\rho}\Delta^{(1)}&=&0, \label{kap1}
 \\
 \chi \dot{\ell}_\bot^{(1)'}+3\dot{\ell}_\bot^{(1)} &=&-2\dot{\Delta}^{(1)}, \label{kap2}
 \\
 \frac{\ell_{||}^{(1)}}{a^2\chi^2}
 &=&-\ddot{\ell}_\bot^{(1)} -3H\dot{\ell}_\bot^{(1)}
 +\frac{\ell_\bot^{(1)}}{a^2\chi^2}+\frac{\ell_\bot^{(1)'}}{a^2\chi},
 \label{kap3}
\end{eqnarray}
where prime denotes derivative with respect to the radial coordinate $\chi$.
By solving Eq.~(\ref{kap1}), we obtain
\begin{eqnarray}
 \Delta^{(1)} (t,\chi)=\frac{D^+(t)}{D^+(t_0)}\Delta^+(\chi)+\frac{D^-(t)}{D^-(t_0)}\Delta^-(\chi), \label{kap4}
\end{eqnarray}
where $\Delta^\pm$ are the growing and decaying modes, respectively. 
Since we consider non-Copernican universes that approach to homogeneous and isotropic universes at early stage, 
we choose $\Delta^-(\chi)=0$. 
By integrating Eq.~(\ref{kap2}) with respect to $\chi$, we obtain
\begin{eqnarray}
 \chi^3\dot{\ell}_\bot^{(1)}(t,\chi)
 =-2\int_0^\chi d\tilde{\chi}\tilde{\chi}^2\dot{\Delta}^{(1)}(t,\tilde{\chi}), \label{kap5}
\end{eqnarray}
where we used the regularity condition of $\dot{\ell}_\bot$ to fix the integral function. 
By integrating Eq.~(\ref{kap5}) with respect to $t$, we obtain
\begin{eqnarray}
 \ell_\bot^{(1)} (t,\chi)
 =-\frac{2}{\chi^3}\int_{t_0}^td\tilde{t}\int_0^\chi d\tilde{\chi}
 \tilde{\chi}^2\dot{\Delta}^{(1)}(\tilde{t},\tilde{\chi}),
 \label{kap6}
\end{eqnarray}
where we set $\ell_\bot^{(1)}(t_0,\chi)=0$ to determine the integral function.
By substituting Eq.~(\ref{kap6}) into Eq.~(\ref{kap3}), we can obtain the perturbation $\ell_{||}^{(1)}(t,\chi)$.

Here, it should be noted that we have one degree of freedom to rescale the radial coordinate $\chi$. 
Under the gauge transformation $\chi \to \chi +\kappa \zeta (\chi)$, 
the metric perturbations transform as
\begin{eqnarray}
 \ell_{||}^{(1)}(t,\chi)\to \ell_{||}^{(1)}(t,\chi)-2\zeta^{'}(\chi),~~~{\rm and}~~~
 \ell_\bot^{(1)}(t,\chi)\to \ell_\bot^{(1)}(t,\chi)-2\frac{\zeta (\chi)}{\chi}.
\end{eqnarray}
We fixed the gauge $\zeta(\chi)$ by setting $\ell_\bot^{(1)}(t_0,\chi)=0$ in Eq.~(\ref{kap6}).



\end{document}